\documentclass{aa}  
\usepackage{graphicx}
\usepackage{txfonts}
\begin{document} 

   \title{Effects of solar evolution on finite acquisition time of Fabry-Perot-Interferometers in high resolution solar physics}


   \author{R.~Schlichenmaier
          \and D.~Pitters
          \and
          J.M.~Borrero
          \and
          M.~Schubert
          }

   \institute{Leibniz-Institut für Sonnenphysik (KIS), 
              Schöneckstr. 6, D-79104 Freiburg\\
              \email{schliche@leibniz-kis.de}
             }

   \date{Received ?; accepted ?}

 
  \abstract
 {The imaging spectro-polarimeter VTF (Visible Tunable Filter) will be operated at the Daniel K. Inouye Solar Telescope (DKIST) in Hawaii. Due to its capability of resolving dynamic fine structure of smaller than 0.05\,arcsec, the finite acquisition time of typically 11\,s  affects the measurement process and potentially causes errors in deduced physical parameters.}
 {We estimate those errors and investigate ways of minimising them.}
 {We mimic the solar surface using a magneto-hydrodynamic simulation with a spatially averaged vertical field strength of 200\,G. We simulate the measurement process scanning through successive wavelength points with a temporal cadence of 1\,s. We synthesise Fe\,I\,617.3\,nm for corresponding snapshots. Besides the classical composition of the line profile, we introduce a novel method in which the intensity in each wavelength point is normalised using the simultaneous continuum intensity, and then multiplied by the temporal mean of the continuum intensity.  Milne-Eddington inversions are used to infer the line-of-sight velocity, $v_{\rm los}$, and the vertical (longitudinal) component of the magnetic field, $B_{\rm los}$.}
   {We quantify systematic errors, defining the temporal average of the simulation during the measurement as the truth. We find that with the classical composition of the line profiles, errors exceed the sensitivity for $v_{\rm los}$ and in filigree regions also for $B_{\rm los}$. The novel method that includes normalisation reduces the measurement errors in all cases. Spatial binning without reducing the acquisition time decreases the measurement error slightly.}
{The  evolutionary time-scale in inter-granular lanes, in particular in areas with magnetic features (filigree), is shorter than the time-scale within granules. Hence depending on the science objective less accumulations could be used for strong magnetic field in inter-granular lanes and more accumulations could be used for the weak granular magnetic fields. As a key result of this investigation, we suggest to include the novel method of normalisation in corresponding data pipelines.
}


\titlerunning{Effects of finite acquisition time in high resolution solar physics}
\authorrunning{Schlichenmaier et al.}

\maketitle
%
\section{Introduction}\label{sec:intro}

The solar photosphere and chromosphere are highly dynamic and exhibit a wealth of magnetic structures and phenomena. The underlying processes operate on small spatial and temporal scales and are a consequence of fundamental interactions between plasma, magnetic fields, and radiation. These interactions leave spectro-polarimetric imprints in absorption and emission lines, which form in photosphere and chromosphere, that can be measured with sophisticated instrumentation. 'Classically', there are two types of instruments: (i) slit-scanning spectrographs and (ii) narrow-band imagers scanning in wavelength through a spectral line. Both types can be equipped with polarimetric modulators. While grating spectrographs have the advantage of spectral integrity, high spectral resolution, and large spectral coverage,  Fabry-Perot-interferometer (FPI) based narrow-band imagers have the advantage of image integrity, large field-of-views, and short cadences. 
New technical developments aim to combine spectral and image integrity together with short cadences in integral field units (IFUs). Presently, two concepts look very promising: Image slicers and micro-lense arrays \citep[see e.g.,][]{2019AdSpR..63.1389J, 2020A&A...634A.131K, 2022arXiv220614294D}.


Here, we investigate the spectral integrity of an FPI-based narrow-band imager that may suffer from scanning in wavelength through a spectral line, resulting in a finite acquisition time, during which the solar scene may change \citep[as already noted in][]{10.1117/1.JATIS.3.4.045002}. We use the Visible Tunable Filter \citep[VTF,~][]{2014SPIE.9147E..0ES} which is planned to be installed in 2023 at the Daniel K. Inouye Solar Telescope \citep[DKIST, ][]{2020SoPh..295..172R}. VTF is an imaging spectro-polarimeter for the wavelength range between 520 and 860 nm. It is based on two FPIs\footnote{At first-light, VTF will be equipped with only one FPI, but the 2nd FPI is manufactured and will be integrated soon after.} which scan the narrow band images in wavelength, i.e., the spectral points of a solar line are not acquired simultaneously, but during a finite acquisition time. In general, the integration time at each wavelength step is determined by the desired signal-to-noise ratio, i.e., polarimetric sensitivity. A default measurement at full spatial resolution of the photospheric magnetic field with VTF in Fe\,I\,617.3\,nm takes 11\,s to reach the desired polarimetric accuracy at 11 wavelength points (see Sect.~\ref{sec:measurement}). VTF is designed to operate at the diffraction limit of DKIST, which has an aperture of 4\,m. I.e., at 520\,nm, scales of 20\,km are resolved on the solar surface. On such small scales, dynamical processes in the solar photosphere potentially lead to spurious signals which spoil the measurement. 

\paragraph{Short time scales observed with GREGOR:}

\begin{figure}
\includegraphics*{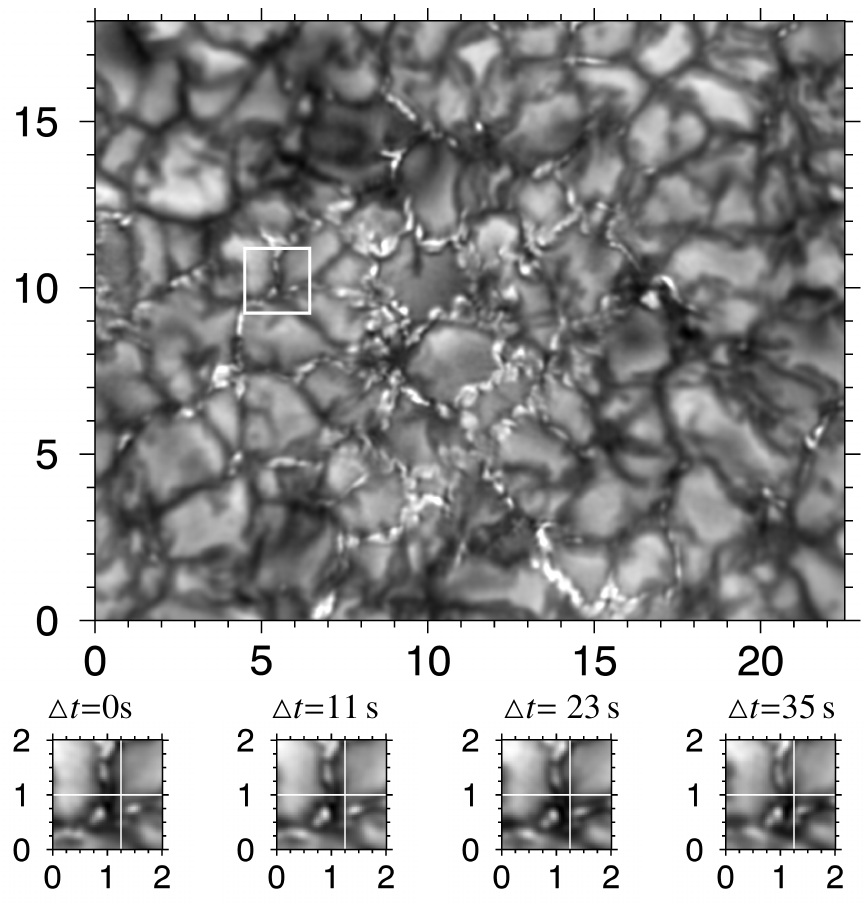}
\caption{\label{fig:gregor}Speckle reconstructed images of a filigree region close to disk center observed with HiFi at GREGOR on June 30, 2019, in G-Band at 430\,nm. Upper large panel shows a large FOV at 08:22:21 UT. The lower four small panel shows four subsequent snapshots $t=$\,0, 11, 23, \& 35\,s of a smaller FOV marked by the white box in large FOV. Tick mark units are in arcsec.
}
\end{figure}

To demonstrate these short time scales in the solar photosphere, in Fig.~\ref{fig:gregor}, we present images from GREGOR \citep{2012AN....333..796S, 2012AN....333..863B} taken with HiFI \citep{Denker_2018} close to disk center on June 30, 2019, in the G-Band. The images are Speckle-reconstructed with KISIP \citep{woeger+al2008} by selecting the 100 best out of 500 images. 

The upper large panel of Fig.~\ref{fig:gregor} shows the observed active filigree region at 08:22:21 UT. Four consecutive snapshots with a time lapse of about 11.5\,s are displayed in the four small panels for the region that is marked with the white box in the large panel. The small panels have a side width of 2 arcsec and contain $67^2$ pixel with a width of 0.03\,arcsec. The solar scene within those 35\,s is clearly not static, and changes on scales of 0.1 arcsec are already visible in consecutive images with a temporal spacing of 11\,s. The evolution is tiny, but clearly visible on small spatial scales. 

In this contribution, we investigate the effects of spurious signals that are expected from temporal evolution during the measurement process. The evolution is mimicked by a realistic magneto-hydrodynamic simulation of a plage region. The VTF measurement process is described in Sect.~\ref{sec:measurement}. The method we apply to mimic the measurement process and how to describe the measurement error is explained in Sect.~\ref{sec:method}. Our Results are presented and discussed in Sect.~\ref{sec:results}. Section \ref{sec:conclusions} presents our conclusion and an important suggestion concerning the data pipeline of Fabry-Perot based spectro-polarimeters.

\section{VTF measurement process}\label{sec:measurement}

The VTF measurement process is described in \citet{2014SPIE.9147E..0ES} and summarised here as follows: The cameras of VTF have a pixel size of 12\,$\mu$m, which corresponds to 0.014\,arcsec and 10 km on the sun. This corresponds to a diffraction-limited spatial resolution of 0.028\,arcsec at a wavelength of 520\,nm. To reach its spectral resolution of $\lambda/\triangle\lambda=100\,000$, the wavelength spacing amounts to 3\,pm at a wavelength of 600\,nm. For our study we use Fe\,I\,617.3\,nm (g=2.5), which has very similar properties as Fe\,I\,630.2\,nm. The VTF cameras are operated with a frame cycle time of 40\,ms, of which 25\,ms are used as exposure time. Simultaneously with each narrow-band image, a broad-band image is recorded in a separate channel of the instrument.

The VTF can be operated in spectroscopic and spectro-polarimetric mode. For both modes, the number of scanning steps, number of accumulations, and the binning size can be adjusted. The VTF Instrument Performance Calculator (IPC, Version 3.4)%
\footnote{\tt  https://www.leibniz-kis.de/en/forschung/ wissenschaftliche-instrumentierung/vtf/ performance-calculator/} was developed to tune these free parameters and to estimate the resulting signal-to-noise ratio (SNR) as well as the time of the measurement. Note that the IPC calculates the SNR for Stokes-I, and does not take into account polarimetric efficiencies. A default spectroscopic measurement with one accumulation at full spatial resolution with an SNR above 200 takes less than one second (=0.88\,s) to measure e.g. Fe\,I\,617.3\,nm with 11 wavelength scan steps (each scan step to tune the etalon takes one camera cycle time of 40ms). In the photosphere, changes of physical parameters on the time scale of one second are not expected within resolution elements as large as 20\,km.

Spectro-polarimetric measurements take more time, on the one hand because four modulation states need to be measured instead of one, and on the other hand because the polarimetric signal in the spectral line is small and requires a large number of photons. To minimise seeing-induced cross talk, besides doing dual-beam polarimetry, the four modulations states are acquired consecutively with four single exposures. To increase the number of photons and reduce the photon noise, consecutive accumulations can be chosen at each wavelength position. With 6 accumulations, a $1\sigma$-noise level of $1/577 =  0.0017$  is reached (with continuum intensity being normalised to 1). To detect a signal, the signal should have a minimum amplitude of $3\sigma$ which corresponds to about 0.005. This limits the minimum magnetic field strength which can be detected. In our case, computing synthetic line profiles for Fe\,I\,617.3\,nm of a quiet Sun model and a spectral resolution of 100\,000, we find that a Stokes-$V$ amplitude of 0.005 is produced by a constant vertical field of  $\sim$ 20\,G at disk center \citep[cf.,][]{2016PhDT.......566S,10.1117/1.JATIS.3.4.045002}. Note that a horizontal homogeneous field needs 175\,G to produce a $Q$-signal of 0.005. With a horizontal field strength of 100\,G, the $Q$-amplitude amounts to 0.0017 \citep[see also][]{2012A&A...547A..89B, 2013A&A...550A..98B}.

\citet{2016SPIE.9908E..4NS} estimate the Doppler shift sensitivity to approximately  80\,m/s, taking into account photon noise and stray light. Note that non-parallelism of the etalon plates were not considered as a source of error in this estimate.

With 6 accumulations, $6\times4=24$ single frames are recorded at each wavelength step, 11 spectral points are measured in 11\,s, i.e., 1\,s per spectral point. This means that a single spectro-polarimetric VTF measurement with 6 accumulation takes 11\,s in total. The task of this contribution is to investigate, whether, based on realistic numerical simulations, one expects dynamic changes in the photosphere within 11\,s, and if yes, how much these changes affect the measurement. Note that many science cases will require a smaller noise level than 0.0017. With 12 accumulations, which take 21.6\,s, VTF reaches a $1\sigma$-noise level of 0.0012. 18 accumulations and 31\,s are needed to reach a noise level of $10^{-3}$. Obviously, measurements with longer scanning times are more affected by the dynamic evolution in the photosphere. But in this contribution, we assume that the measurement process takes 11 seconds for 11 scan steps, i.e., we imitate the case of 6 accumulations.

\section{Methods}\label{sec:method}

To mimic the measurement process of VTF, we take consecutive snapshots of a realistic magnetohydrodynamic simulation of a solar plage region (cf.~Sect.~\ref{sec:muram}), synthesise Stokes profiles, and adapt them to the VTF spectral resolution and acquisition procedure (cf. Sect.~\ref{sec:firtez}). In Sect.~\ref{sec:normalisation} we introduce an alternative way to construct the line profiles, normalising each wavelength point to the local continuum intensity. 
As a result, we produce different sets of Stokes profiles from the 11 time steps of the simulation. In Sect.~\ref{sec:vfisv} we describe how these different sets of profiles are inverted to retrieve the physical parameters of line-of-sight velocity and and vertical component of the magnetic field strength.

To quantify the error of the measurement process, we need to define a reference or {\it true} map. As {\it true} maps, we use the maps that result from the inversion of profiles that are synthesised from time-averaged simulation boxes (over all eleven snapshots). I.e., these profiles are not affected by the finite acquisition time and therefore serve as reference.

Finally, all different sets of maps are compared in Sect.~\ref{sec:results}.  By comparing physical parameters between inverted maps from constructed profiles and the reference maps, we can estimate the error which is introduced by the finite acquisition time.

\subsection{Solar plage simulation with MURAM}\label{sec:muram}

\begin{figure}
\includegraphics*{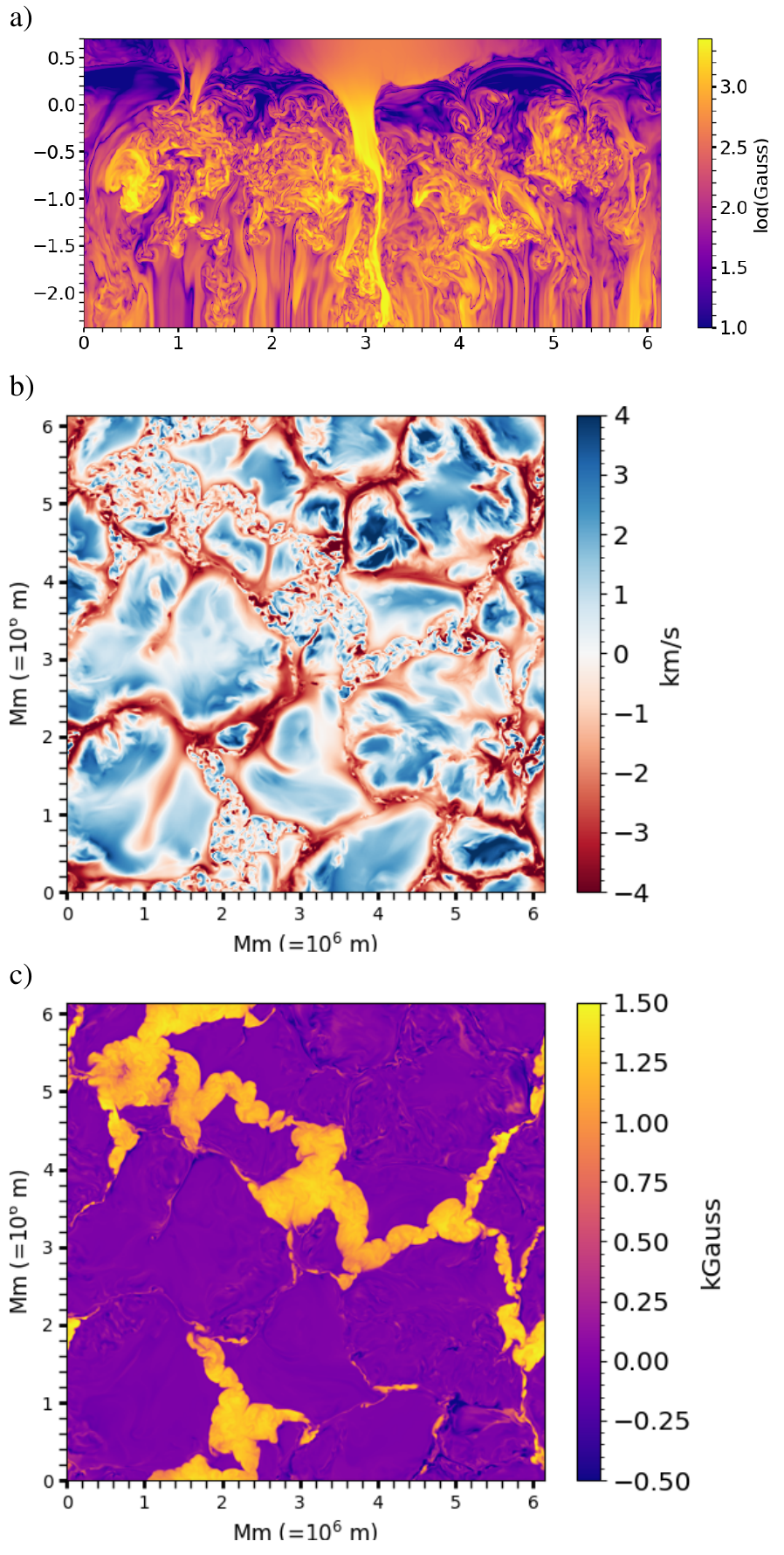}
\caption{\label{fig:muram} MURAM simulation snapshot. a) vertical cut through numerical box showing the modulus of the magnetic field strength in logarithmic scale, clipped between 10 and 2500\,G, with the color bar in logarithmic scale. b)  Horizontal cut at mean temperature of 5560\,K of vertical velocity in km/s with positive values pointing upwards (being blue-shifted to observer). c) Same as b) for vertical component of magnetic field strength in kG.}
\end{figure}

As a typical target for VTF observations we take a plage region with a spatially averaged vertical magnetic strength of 200\,G. Such a simulation was performed with MURAM by Matthias Rempel \citep[private communications, and see][]{2014ApJ...789..132R}. With a grid cell size of 8x8x8 km$^3$, the simulation reaches the spatial resolution of VTF@DKIST. The simulation box has 768x768x384 cells. The box reaches from the upper photosphere into the solar interior. The upper end of the box lies 704\,km above the average $\tau=1$ level. As the lateral box sides have periodic boundary conditions, the magnetic flux through each depth layer is the same and constant in time.

The internal numerical time step of the simulation is around 0.1\,s. We limit our analysis to snapshots with a temporal cadence of 1\,s, i.e. 10 numerical time steps. Changes within 10 time steps can be considered small for our purpose, but contribute to an increase of the measurement errors. For this work, we assume that it is sufficient to have a snapshot sequence of 1\,s. This is the same cadence that VTF needs for each wavelength step when using 6 accumulations (SNR$=577$). 

In Fig.~\ref{fig:muram}a, we display a vertical cut of the absolute magnetic field strength of the computational box. It displays a vertical strong flux concentration in the middle of the box at $x=3$\,Mm. There, the local field strength exceeds 3\,kG. For the horizontal layer at which the spatially averaged temperature is 5560\,K, we show the vertical components of the velocity and the vertical magnetic field strength in Figs.~\ref{fig:muram}b \& c. This horizontal layer roughly corresponds to the formation height of Fe\,I 630.25\,nm such that these maps can be compared with the maps that we retrieve after line synthesis, convolution with the VTF spectral resolution, resampling, and Milne-Eddington inversion.

\subsection{Synthetic line profiles with FIRTEZ adapted to VTF}\label{sec:firtez}

From a number of suitable photospheric absorption lines we chose to analyse the Fe\,I\,617.3 nm.\footnote{This line has very similar properties as the iron lines at 630\,nm and at 525.0\,nm \citep{2003ASPC..307..131G}. Initially, the intention of this work was to compare VTF measurements with HMI onboard SDO. The focus changed, and we continued to use Fe\,I\,617.3\,nm. Fe\,I\,617.3\,nm has the advantage of being surrounded by a clean continuum.} VTF will use Fe\,I 630.25\,nm for the first observations as long as only one etalon is available, but Fe\,I\,617.3\,nm will be available as soon as the second etalon (Fabry-Perot interferometer) is in place.

To synthesise Stokes $I$, $Q$, $U$ $\&\,V$ of Fe\,I\,617.3 nm along vertical line-of-sights in the MURAM box, we use FIRTEZ-dz \citep{2019A&A...629A..24P}. FIRTEZ is chosen because it operates on the geometrical scale, which is intrinsic to the simulation box. As the conversion into optical depth along the line-of-sight is not needed, the computation time for the line synthesis is very short. We calculate the set of Stokes profiles for each horizontal pixel yielding maps with 768 by 768 pixel.

For the line synthesis the spectral resolution is assumed to be infinity and is only limited by the spectral sampling. We use a spectral sampling of 0.4\,pm, and compute the range from -40\,pm until +40\,pm.

\begin{figure}
    \centering
    \includegraphics[width=0.5\textwidth-0.5\columnsep]{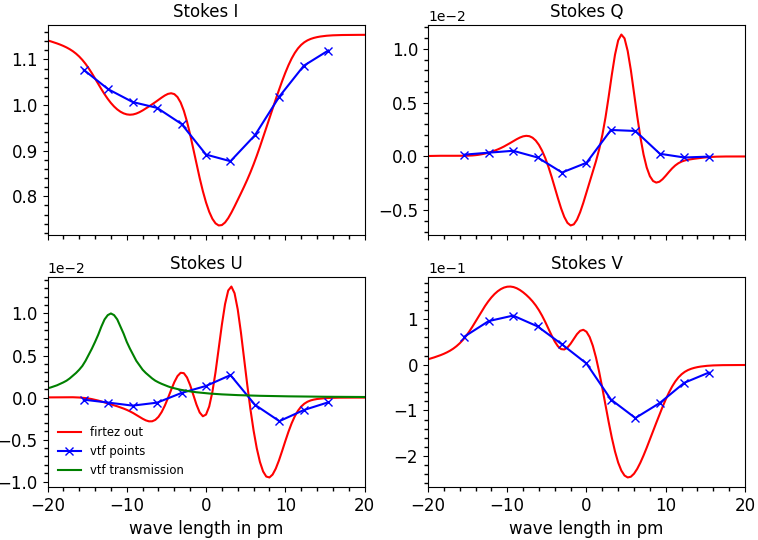}
    \caption{\label{fig:convolution} Stokes I, Q, U, and V profiles from upper left to lower right panels, respectively. Red line denotes synthesised profiles with intrinsic sampling of 0.4\,pm. Green line in the lower left panel sketches the shape of the VTF transmission curve. Blue crosses and lines mimic 11 VTF measurement point as a result of convolving the synthetic profiles with the VTF transmission curve.  }
\end{figure}

\paragraph{Spectral convolution and spectral sampling:} The spectral resolution of VTF follows from the properties of the first Fabry-Perot interferometer. The spectral transmission profile \citep[see e.g., Sect.~3.4.4 in][]{2002tsai.book.....S} can be computed for small angles of incidence from its reflectance, $R=0.95$, and its cavity separation, $d=0.55$\,mm \citep[][]{Kentischer+al2012, 2014SPIE.9147E..0ES, sigwarth2017}. A prefilter is assumed to suppress the side peaks and to transmit only the central peak.
The transmission profile of the central peak corresponds to a spectral resolution of 100\,000 and has a full width at half maximum of 5.7\,pm. All FIRTEZ profiles are convolved with this transmission profile and resampled to the VTF spectral step width of $\triangle\lambda=3.15$\,pm. A set of sample profiles are depicted in Fig.~\ref{fig:convolution}: The synthetic lines of Stokes $I$, $Q$, $U$, and $V$ of a sample pixel are drawn in red. The blue lines display the same profiles after accounting for the spectral resolution of VTF, and the blue crosses mark the spectral sampling of VTF. To sketch the shape of the tuneable VTF transmission profile, it is plotted as a green line in the Stokes $U$ plot.

Thus, in the end we have Stokes profiles (at the spectral resolution and sampling of VTF) for a set of 11 snapshots, which have a temporal cadence of 1\,s.

\paragraph{Temporal sampling:} To mimic the VTF line scan, we compose the Stokes profiles with one wavelength step per consecutive snapshot, i.e., we produce a set of profiles with 11 ($t=0,1,\ldots, 10$\,s) wavelength points.
We do this in two modes: straight and rabbit. In the {\it straight} mode the wavelength step increases monotonically by one: $\lambda_{i+1} = \lambda_i + \triangle\lambda$. In the {\it rabbit} mode we jump from blue wing to red wing, successively approaching the line core: $\lambda_{i+1} = \lambda_i + 2*(\lambda_{ic}-\lambda_i)$ and $\lambda_{i+2}=\lambda_{i}+\triangle\lambda$, with $\lambda_{ic}$ being the wavelength step closest to the spatially averaged line core. The idea of the rabbit mode is to sample both line wings as close in time as possible. Herewith, one expects that the Doppler shift error from the finite acquisition time is minimised.

\paragraph{Seeing and solar evolution:} The rabbit mode was devised for VTF to also minimise the effects of seeing. The seeing is partially corrected by the adaptive optics system. The resulting degree of image correction varies on scales of seconds and shorter. With the rabbit mode, one attempts to record both line wings with similar seeing. However, the effects of seeing are not analysed in this work. We ignore seeing effects, and only consider the effects of solar evolution during the line scan.

\begin{figure}
    \centering
    \includegraphics*[width=0.5\textwidth-0.5\columnsep, bb=50 0 600 400]{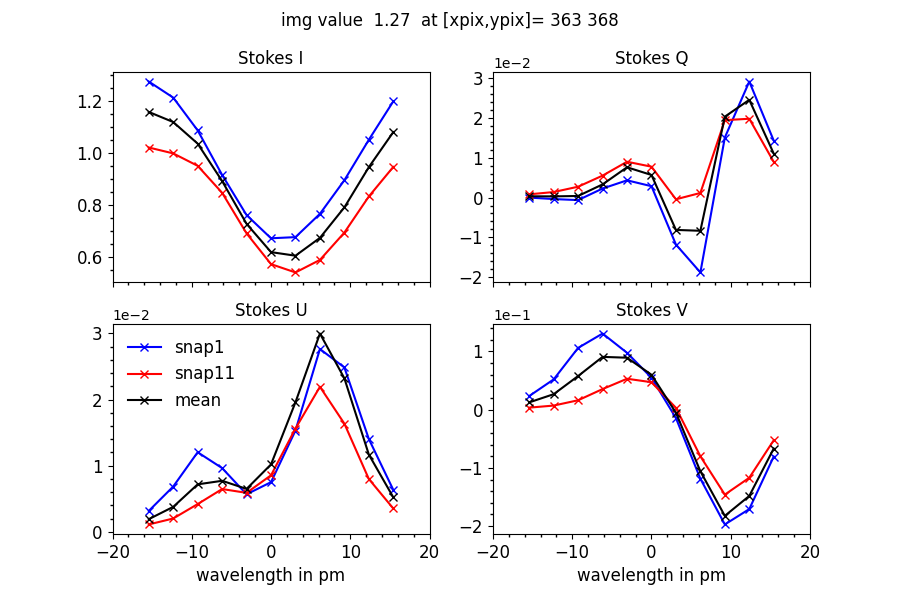}
    \caption{ \label{fig:temporal_change}
 For an example pixel we plot the Stokes profiles for $t=0\,$s (blue) and $t=10\,$s (red). The black curve shows the profile from a temporal average of all 11 snapshots.}
   \end{figure}

\paragraph{Reference profile and temporal changes:} In order quantify the systematic measurement errors, we need to define a reference. We use profiles from temporally averaged snapshots for each pixel as the reference, i.e. we average all 11 snapshots and perform the line synthesis on the averaged box. Alternatively, one can average the synthesised profiles from the 11 snapshots and do the inversion on the time-averaged profiles. We checked that both approaches lead to identical results within the  noise, which is on the same scale as the difference between two consecutive snapshot maps.

In Fig.~\ref{fig:temporal_change} we illustrate the temporal change of the profiles. We select a sample pixel and plot the Stokes profiles retrieved from the first snapshot 1 (blue line), from the last snapshot 11 (red line), and the temporally-averaged profiles from all 11 snapshots (black line). In the upper left panel, it is seen that the intensity of the left-most wavelength point drops from some 1.28 to some 1.02, i.e., by more than 25\% during the measurement process across 11 wavelength points. 

\subsection{Profile composition with normalisation}\label{sec:normalisation}

\begin{figure}
\begin{center}
\includegraphics{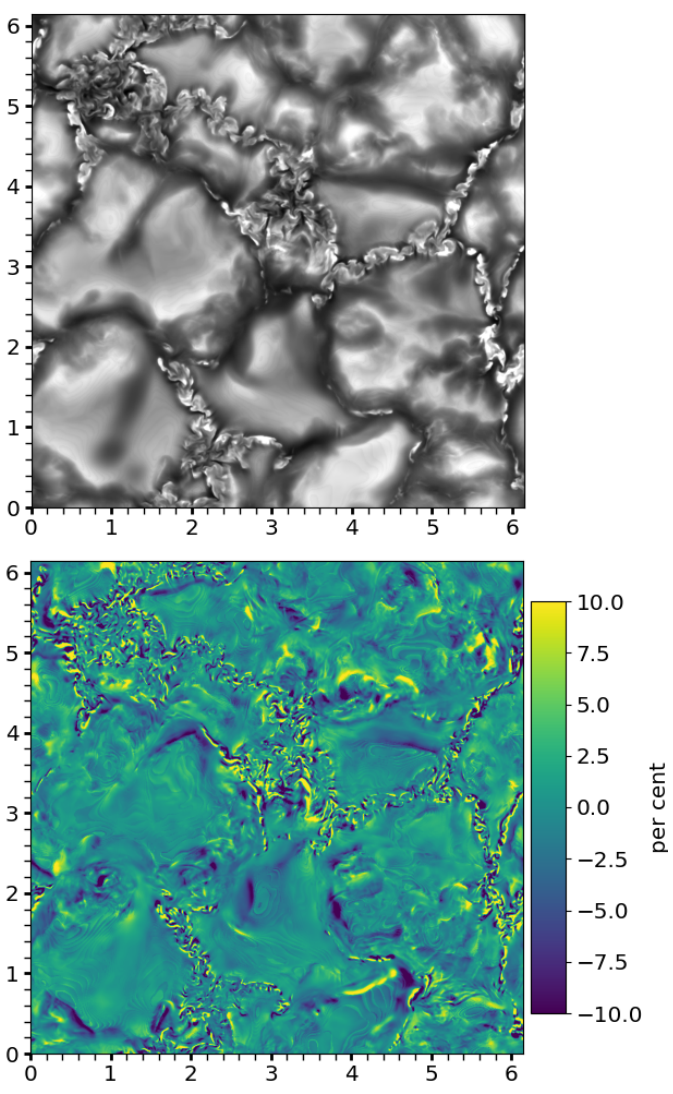}
\end{center}
\caption{\label{fig:cont_diff}
In the upper panel the continuum image with  intensities between 0.6 and 1.45 for the first snapshot is displayed. Average intensity is normalised to one. The lower panel displays the difference between the continuum images of snapshot 1 and of snapshot 11. The difference map is clipped at differences of $\pm10$\,\%.
}
\end{figure}

Viewing animations of the snapshots, it becomes apparent that magnetic concentrations, which are visible as bright points in the inter-granular lanes, are shuffled around by the granular flow field. As a consequence, the continuum intensity of an inter-granular pixel can change significantly during the wavelength scan. This is demonstrated in Fig.~\ref{fig:cont_diff}, in which we display the normalised continuum intensity of the first snapshot and the difference map between the first (snap1) and last (snap11) snapshot. The difference map is clipped at absolute differences of 0.1. We compute that in 13\% of all pixels the intensity difference is larger than $0.05$, and in 4\% of all pixels the difference is larger than 0.1. The standard deviation amounts to 0.034. In comparison, the standard deviation of intensity differences between first (snap1) and second (snap2) snapshot amounts to 0.006.

If the continuum intensity, for example, decreases continuously during a straight wavelength scan, an analysis of the corresponding composed Stokes-$I$ profile would yield different intensity levels of the blue and red continua. This would be interpreted as an additional line absorption in the red wing, i.e., as an enhanced red-shift. This example suggests that it might be of advantage to normalise the line absorption to the continuum. Since we have the full simulated profiles at each time step (= wavelength step), it is possible to do this. Obviously, this is not straight forward for the VTF measurement. However, as VTF is operated with a simultaneous broad-band channel, this normalisation could be performed using the broad-band intensity as it varies during the line scan at a given spatial pixel.

\begin{figure}
    \centering
    \includegraphics*{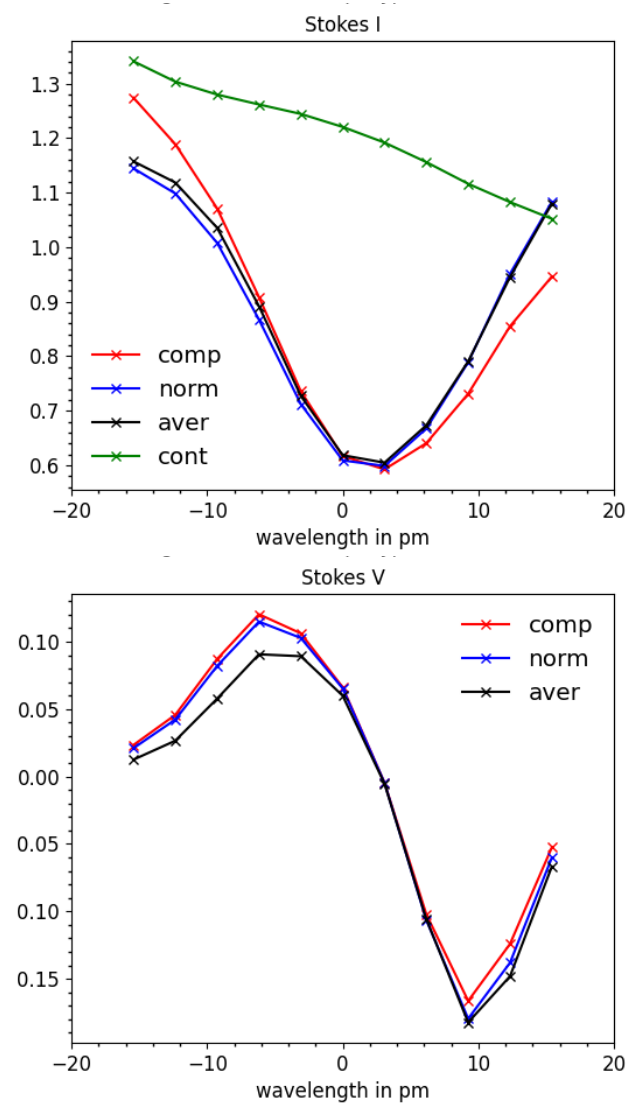}
    \caption{\label{fig:composition} Exemplary comparison of Stokes-$I$ and $V$ profiles in upper and lower panel, respectively: straight composed (red) simulating the VTF measurement, straight normalised (blue), and time-average. The green line in upper panel denotes the continuum intensity at each wavelength (=time) step.}
\end{figure}

In Fig.~\ref{fig:composition} we illustrate how the straight profile is normalised, and compare the different Stokes-$I$ profile types for the same spatial pixel as in Fig.~\ref{fig:temporal_change}. The black line displays the time-averaged profile. The red line displays the straight VTF measurement mode. The rabbit mode behaves similar as the straight profiles and is not displayed. The green line denotes the continuum intensity variation during the measurement process. From these continuum points, the temporal mean is calculated. This mean value, $\langle I({\rm cont}_{\rm i}) \rangle_{i=1,\ldots,11}$, is used to normalise each wavelength point of the straight profile (red line):
\begin{equation} \label{eq:norm}
    I_{\rm normalised}(\lambda_{i}) = \frac{ I(\lambda_{i}) }{I({\rm cont}_{i})} \cdot 
     \langle I({\rm cont}_{i}) \rangle_{i=1,\ldots,11}
\end{equation}
This normalised profile is drawn as blue line in Fig.~\ref{fig:composition}. It is seen that the normalised straight profile becomes very similar to the time-averaged profile. Therefore, one expects that the normalised profiles leads to physical parameters that are closer to the reference (time-averaged) parameters. For pixels in which the continuum intensity is constant in time, the straight, normalised, and time-averaged profiles are identical.

\subsection{Milne-Eddington inversion with VFISV}\label{sec:vfisv}

To compare the differently composed Stokes profiles, we perform Milne-Eddington inversions and compare inverted maps of the line-of-sight velocities, $v_{\rm los}$, and of the vertical (line-of-sight) component of the magnetic field strength, $B_{\rm los}$, for the different profile types. The Milne-Eddington approximation assumes that $v_{\rm los}$ and $B$ as well as $B_{\rm los}$ are constant along the line-of-sight across the solar photosphere. Inspecting the numerical simulation (see, e.g., the vertical cut of $B_{\rm los}$ in Fig.~\ref{sec:muram}), this is clearly not the case, and these quantities change significantly along the line-of-sight. However, in order not to introduce additional error sources due to sophisticated inversions, we limit our analysis to Milne-Eddington inversions.

As inversion code, we use ``\underline{V}ery \underline{F}ast \underline{I}nversion of the \underline{S}tokes \underline{V}ector'' \citep[VFISV, ][]{2011SoPh..273..267B}. This code was originally devised to invert data from the Helioseismic and Magnetic Imager (HMI) onboard the Solar Dynamic Observatory (SDO), and was later adapted to data acquired with slit spectrographs\footnote{\texttt{gitlab.leibniz-kis.de/borrero/vfisv\_spec}}as, e.g., for GRIS@GREGOR \citep[][]{2012AN....333..872C}, and to data acquired with Fabry-Perot systems\footnote{\texttt{gitlab.leibniz-kis.de/borrero/vfisv\_fpi}}. In the latter version, the VTF transmission file can be selected in such a way that it is taken into account in the inversion process. Note that we use the identical transmission profile in the convolution with the synthetic profiles in Sect.~\ref{sec:firtez}.
 As free parameters, we use $\eta_0$, field inclination $\gamma$, field azimuth $\phi$, damping $a$, doppler width $\triangle\lambda_{\rm D}$, field strength $B$, line-of-sight velocity $v_{\rm los}$, source function continuum, and the source function gradient. The magnetic filling factor is put to unity and not used as free parameter.

\section{Results and discussions}\label{sec:results}

\subsection{Inverted maps for reference (time-averaged) profiles}\label{sec:ref_maps}
\begin{figure}
    \centering
    \includegraphics*{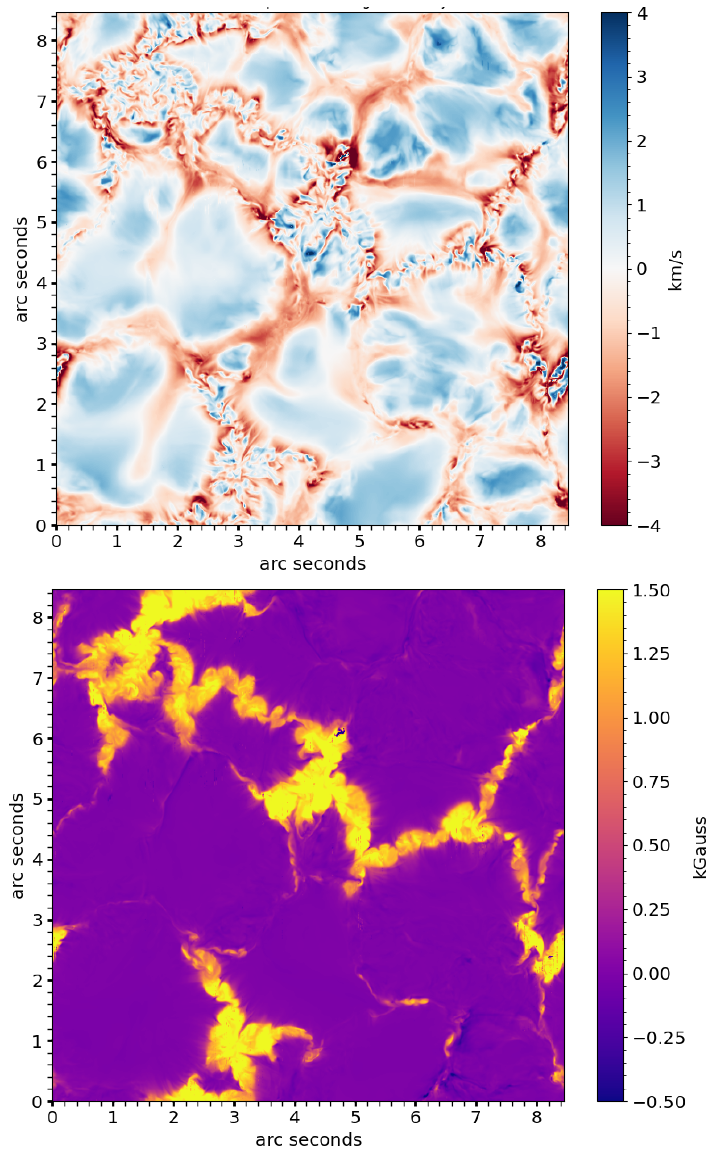}
    \caption{\label{fig:ref_map} Maps for line-of-sight velocity, $v_{\rm los}$ (upper panel), and vertical component of the magnetic field strength, $B_{\rm los}$ (lower panel) inverted with VFISV from reference (time-averaged) Stokes profiles. As these maps are simulated observables, we use arcsec instead of km for the spatial dimension.}
\end{figure}

\begin{figure}
    \includegraphics*{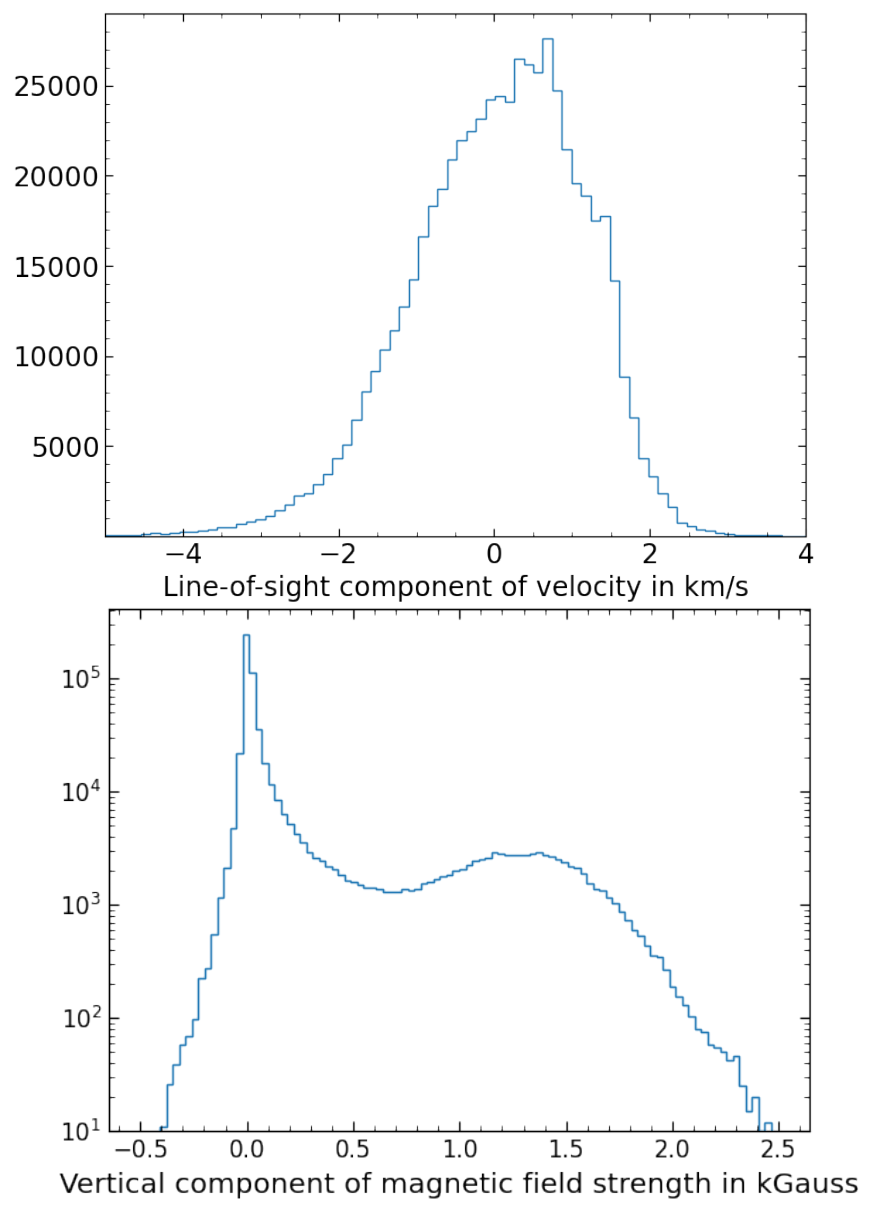}
    \caption{\label{fig:ref_hist} Histograms for maps of line-of-sight velocity with a bin size of 115 m/s (upper panel), and vertical component of the magnetic field strength with a bin size of 30\,G (lower panel) for the inverted maps of the reference (time-averaged) profiles.}
\end{figure}

Inverted maps for $v_{\rm los}$ and $B_{\rm los}\!=\!B\cdot \cos(\gamma)$ of the reference (time-averaged) profiles are displayed in Fig.~\ref{fig:ref_map}. These maps correspond to the average along the line-of-sight, but for consistency, they can be compared to the horizontal geometrical cuts at the mean temperature of 5560\,K in Fig.~\ref{fig:muram} b) and c). In both figures, we use the same color code and clipping values. They are obtained after line synthesis with FIRTEZ on the time averaged numerical box, convolution with the instrument transmission profile (Sect.~\ref{sec:firtez}), and VFISV inversion (Sect.~\ref{sec:vfisv}). 

These maps reflect the ideal theoretical expectation of what VTF@DKIST will measure in a plage region with an average vertical magnetic field strength of 200\,G. It is therefore of interest to plot the histogram distribution for $v_{\rm los}$ and $B_{\rm los}$ in Fig.~\ref{fig:ref_hist}. Both distributions are non-symmetric. The up-flow velocities do not reach as high values as the down-flow velocities, but occupy a larger area for small absolute values. Opposite polarity in the magnetic field is present and reaches values of up to -400\,G. The distribution of the positive vertical component of the magnetic field peaks at some 1300\,G and $B_{\rm los}$ reaches values of up to 2.4\,kG.

\subsection{The average of the vertical magnetic field strength}

The horizontal average vertical magnetic field strength (magnetic flux through a horizontal cut) in the simulation box is constant at all depth layers and at all times due to the periodic boundary conditions on the sides of the box. It was chosen to be 200\,G. This number can be compared with maps of $B_{\rm los}$ for the various types of profiles: Averaging the $B_{\rm los}$ across the inverted maps we obtain $\langle B_{\rm los}\rangle\!=\!208$\,G for the reference profiles, 203\,G for the straight VTF profiles, 206\,G for the rabbit VTF profiles, 208\,G for the straight normalised VTF profiles, and 208\,G for profiles of each snap shot. 

The difference of 8\,G between the numerical box value and the reference map is ascribed to the different geometrical formation heights. We surmise that strong field concentrations are associated with smaller densities, due to magnetic evacuation (magnetic pressure). Hence, in areas of magnetic field concentration, the opacity is reduced and the line forms in deeper geometrical layers in which the field strength is stronger. This effect could explain that the average value $\langle B_{\rm los}\rangle $ in the inverted reference map is larger by 8\,G than in the numerical box at constant geometrical height. In any case, the difference is small and not significant for our study. 

More interesting for our study is the finding that normalised straight VTF profiles yield the same $\langle B_{\rm los}\rangle$ as the reference profiles, while the straight VTF profiles differ by 5\,G. This difference is small, but confirms the expectation (Fig.~\ref{fig:composition} in Sect.~\ref{sec:normalisation}) that the normalised VTF profiles are closer to the reference (time averaged) profiles.

\begin{figure}
\centering
\includegraphics{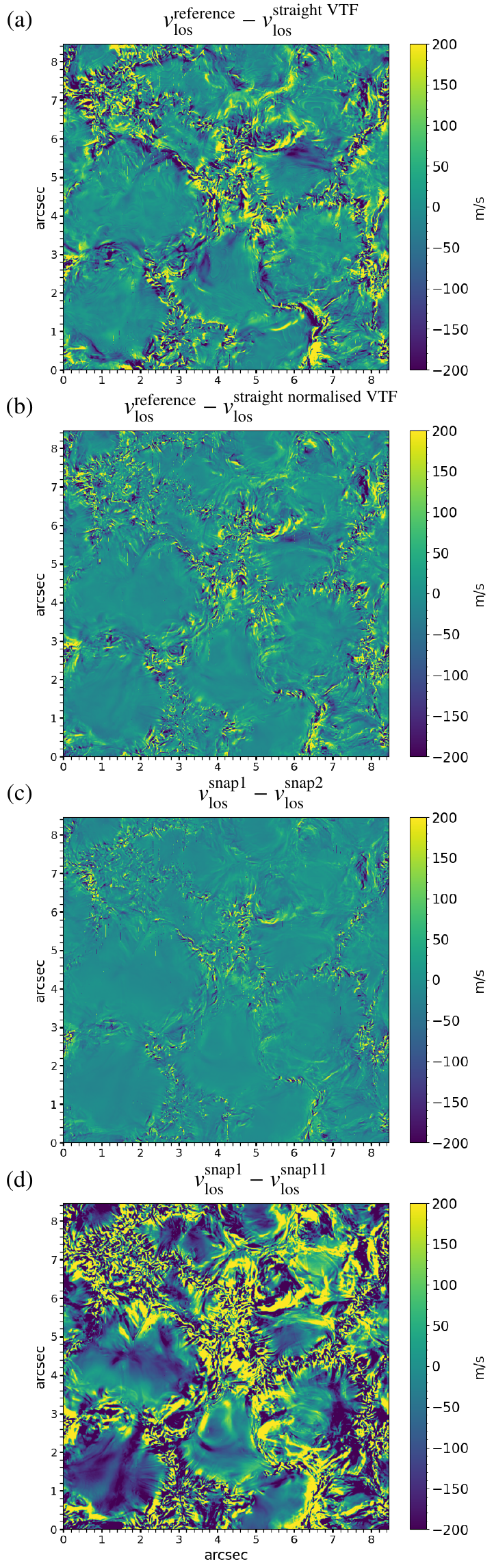}
\caption{\label{fig:diff_vlos} Difference maps for $v_{\rm vlos}$ in m/s. From top to bottom: Difference to reference map of straight VTF map (a); same for straight normalised VTF map (b); Difference of first snapshot 'snap1' to second snapshot 'snap2' (c); same for eleventh snapshot 'snap11' (d).} 
\end{figure}

\begin{figure}
\centering
\includegraphics{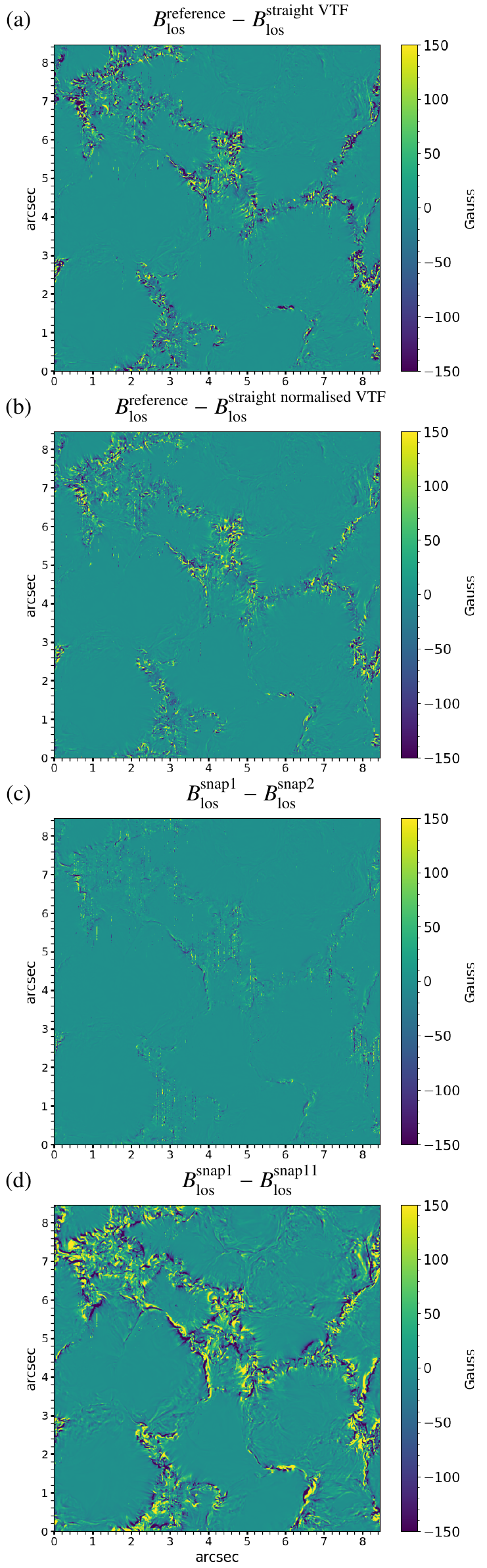}
\caption{\label{fig:diff_Bver} Same as Fig.~\ref{fig:diff_vlos} for $B_{\rm los}$ in Gauss.}
\end{figure}

\subsection{Oscillation of the numerical box}

As the sun, the MURAM box exhibits 5-minute oscillations that are visible in maps of $v_{\rm los}$. Therefore the average line-of-sight velocity is not constant. The particular snapshots that we analyse change from a mean velocity of 41\,m/s in the first snapshot, by some 2\,m/s per second, to 18\,m/s in the eleventh snapshot. Hence, the spatially averaged $v_{\rm los}$ changes by 21\,m/s during the analysed VTF measurement process.
 
\subsection{Difference maps for maps of $v_{\rm los}$ \& $B_{\rm los}$}

As explained in Sect.~\ref{sec:intro}, the purpose of this study is to quantify the measurement error due to a finite measurement time of 11\,s. To determine the error we compare simulated VTF maps to the reference (time-averaged) maps by computing difference maps.  In the upper two panels of Figs.~\ref{fig:diff_vlos} and \ref{fig:diff_Bver}, we display the difference maps from the inversion of straight VTF profiles and the straight normalised VTF profiles with the reference maps. For $v_{\rm los}$, the differences are clipped at $\pm 200$\,m/s, and for $B_{\rm los}$ at $\pm150$\,G. By visual inspection it is seen that the differences are smaller for the straight normalised profiles if compared to the straight VTF profiles. The improvement of the normalisation is clearly seen in $v_{\rm los}$ and less prominent but still visible in $B_{\rm los}$. These differences are quantified in the next subsection.

The lower two panels in Figs.~\ref{fig:diff_vlos} and \ref{fig:diff_Bver} illustrate the solar evolution during the measurement process by displaying the difference maps between first and second snapshot (separated by one second, snap1 \& snap2) and between first and eleventh snapshot (separated by ten seconds, snap1 \& snap11).  As expected the difference between snap1 and snap11 are much larger than between snap1 and snap2. The differences between snap1 and snap2 can have two causes: solar evolution and/or sensitivity of the inversion process. As the solar evolution on a time scale of one second is expected to be insignificant, we ascribe these differences mostly to the sensitivity of the inversion process which is limited by the spectral resolution and the Milne-Eddington approximation.

\subsection{Error quantification}

\begin{figure}
    \includegraphics{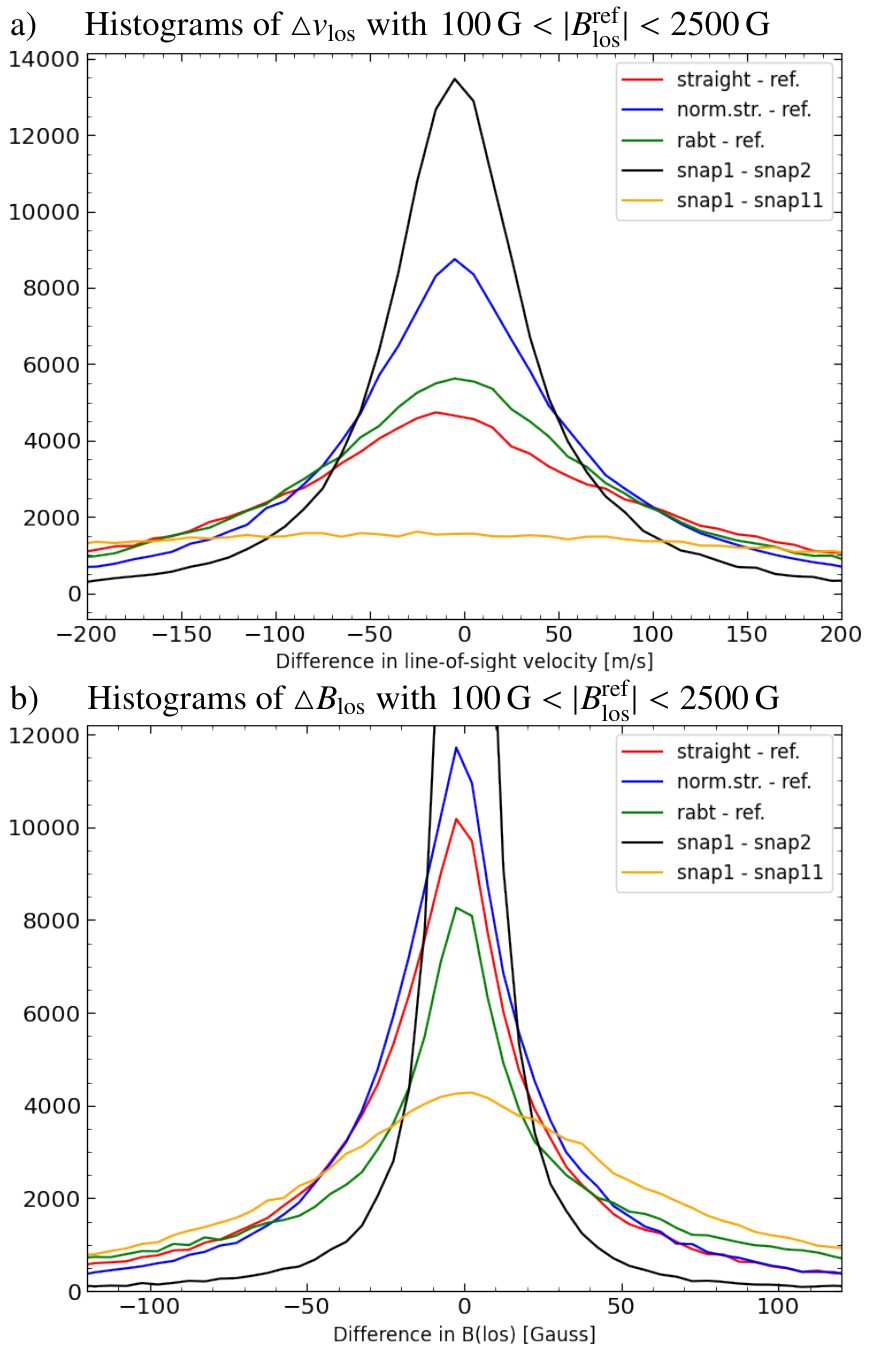}
    \caption{\label{fig:histograms} Histograms of difference maps for line-of-sight velocity. $\triangle v_{\rm los}$ in upper panel (a) and the vertical component of the magnetic field strength, $\triangle B_{\rm los}$, in lower panel (b). The histograms includes pixel that have $100\,{\rm G} < |B_{\rm los}^{\rm ref}| < 2500\,{\rm G}$. This corresponds to 25\% of all pixels.
The differences are taken for different cases: Red line $\rightarrow$ straight VTF and reference map; blue line $\rightarrow$ normalised straight VTF and reference map; green line $\rightarrow$ rabbit VTF and reference map; black line $\rightarrow$ first and second snapshot; Yellow line $\rightarrow$ first and last (eleventh) snapshot.}
\end{figure}

In order to quantify the visual impression of the difference maps, we compute histograms and plot them in Fig.~\ref{fig:histograms}. As the deviations are largest in the inter-granular lanes and in the filigree (magnetic) regions, we consider only pixels in which the vertical line-of-sight field strength of the reference map exceeds 100\,G. We ignore 67 pixels with $B_{\rm los} > 2500$\,G as outliers. The histograms then include 25\% of the pixels, i.e., 146\,529 out of 589\,824 pixels. The different line colours correspond to the different cases:
\begin{itemize} 
\item ('straight VTF' $-$ 'reference') $\rightarrow$ red line
\item ('normalised straight VTF' - 'reference') $\rightarrow$ blue line
\item ('rabbit VTF' - 'reference') $\rightarrow$ green line 
\item ('snapshot 1' - 'snapshot 2') $\rightarrow$ black line
\item ('snapshot 1' - 'snapshot 11') $\rightarrow$ yellow line
\end{itemize}

The histograms of the $v_{\rm los}$ maps are plotted with a bin width of 10\,m/s in Fig.~\ref{fig:histograms}a. The distributions peak close to vanishing differences. The peak height and the width are a quantitative measure for the measurement error: the larger the peak and the smaller the width, the smaller the measurement error. The visual impression of Fig.~\ref{fig:diff_vlos} is reflected in the distributions in terms of peak height and width: 'snap1 - snap2' has the largest peak and the smallest width, i.e., the smallest errors, and 'snap1 - snap11' has the smallest peak and the largest width, i.e. the largest errors. The histogram 'normalised straight - reference' indicates smaller errors than 'straight - reference'. It is also seen that the performance of the 'rabbit' mode (green line) is not better than the 'straight' mode (red line).

The histograms of the $B_{\rm los}$ maps are plotted with a bin width of 5\,G in Fig.~\ref{fig:histograms}b. Again, the black line ('snap1 - snap2') has the largest peak (at 30521). The ordinate is clipped at a value of 12\,000 to increase the visibility of the blue line ('normalised straight') and the red line ('straight'). The latter distributions peak at 11\,716 and 10\,177, respectively, i.e., again, the 'normalised straight' mode performs better than the 'straight' mode. The green line ('rabbit' mode) is worse and has a peak at 8\,263. As expected the differences between snap1 and snap11 are the largest such that the central peak of smallest differences only occurs 4\,276 times.

The histogram distributions in Fig.~\ref{fig:histograms} are not Gaussian. In a Gaussian distribution, 68.27\% of all values would lie within one standard deviation, $1\sigma$, i.e. the full width including 68.27\% corresponds to $2\cdot\sigma$, and at that level the peak value is reduced by a factor of 1.65. Analogously, we determine the width, $w_{\rm full}$, of the distribution that include 68.27\% of all values, and then quantify the error as $1\sigma := w_{\rm full}/2$. 

\begin{table}
\caption{\label{tab:vlosmaps} Half of full distribution width, $w_{\rm full}/2$, corresponding to $1\sigma$, for difference maps of $v_{\rm los}$ in m/s. The bin width is set to 10 m/s.}
\begin{tabular}{c|c|c|c|c}
$\triangle v_{\rm los}$  & $w_{\rm full}/2$  & $w_{\rm full}/2$  & $w_{\rm full}/2$   & $w_{\rm full}/2$  \\ 
unit                                    & [m/s]                   & [m/s]                   & [m/s]                     & [m/s]     \\ \hline
'straight' 	                         &  67                   & 46                     & 181                      & 201    \\
{\!'norm.~str.'\!}                   & 36                    & 25                     & 88                    &  94     \\
'rabt'  	                         & 55                   & 39                    & 146                   & 159    \\
's1-s2' 	                         &  18                     & 12                      & 49                   & 55    \\
's1-s11' 	                         & 195                    & 142                   &  432                  & 470 \\\hline
px  fract.                            & {100\%}              & {75 \%}           & {25 \%}               & {20\%} \\
$\!|B_{\rm los}|$ mask\!   &  {\!all pixel}\!    & {$\!0\!-\!100$G\! } & {$\!0.1\!-\!2.5$kG\!}  & {$\!0.2\!-\!2.5$kG\!} \\
\end{tabular}
\end{table}

\begin{table}
\caption{\label{tab:Bvermaps} Same as Tab.~\ref{tab:vlosmaps} for deviations of $B_{\rm  los}$ in G. For the calculation of $w_{\rm full}$, the bin width is set to 1 Gauss.}
\begin{tabular}{c|c|c|c|c}
$\triangle B_{\rm los}$  & $w_{\rm full}/2$  & $w_{\rm full}/2$  & $w_{\rm full}/2$   & $w_{\rm full}/2$  \\ 
unit                                    & [G]                   & [G]                & [G]                     & [G]     \\ \hline
'straight' 	                         &  4                   & 2                   & 48                      & 61    \\
{\!'norm.~str.'\!}                  & 3                    & 1                   & 36                      & 44     \\
'rabt'  	                         & 3                    & 1                   & 84                       & 109    \\
's1-s2' 	                         &  2                     & 1                   & 10                     & 12    \\
's1-s11' 	                         & 16                    & 7                 &  91                      & 105 \\\hline
px  fract.                            & {100\%}           & {75 \%}          & {25 \%}             & {20\%} \\
$\!|B_{\rm los}|$ mask\!   &  {\!all pixel}\!    & {$\!0\!-\!100$G\! } & {$\!0.1\!-\!2.5$kG\!}  & {$\!0.2\!-\!2.5$kG\!} \\
\end{tabular}
\end{table}

\begin{table}
\caption{\label{tab:binning} Half of full distribution width, $w_{\rm full}/2$, corresponding to $1\sigma$, for four different binning cases using the $B_{\rm los}$-mask $0.1-2.5$\,kG. Values are given in m/s. The bin width is set to 10 m/s.}
\begin{tabular}{c|c|c|c|c}
binning & 1 by 1  & 2 by 2  & 3 by 3 & 4 by 4   \\  \hline
'straight' 	                         &   181                &    169                  &      159        & 148             \\
{\!'norm.~str.'\!}                  &    88                 &      77                &            68       & 61 \\
'rabt'  	                         &    146               &     134                 &      121       &    114   \\
's1-s2' 	                         &    49                 &      44              &             39    &  35 \\
's1-s11' 	                         &     432              &       409             &          378 &      349   \\  \hline
\end{tabular}
\end{table}

These $w_{\rm full}/2$-values are given in Tabs.~\ref{tab:vlosmaps} \& \ref{tab:Bvermaps} for $v_{\rm los}$ and $B_{\rm los}$, respectively. We determine the distribution widths for four different mask criteria: all pixels (first column), pixels with $0<|B_{\rm los}|<0.1$\,kG (2nd column, 75\%), pixels with $0.1<|B_{\rm los}|<2.5$\,kG (3rd column, 25\%), pixels with $0.2<|B_{\rm los}|<2.5$\,kG (4th column, 20\%).

\subsubsection{Errors in $v_{\rm los}$ maps}

The Doppler shift sensitivity of VTF was estimated (see Sect.~\ref{sec:measurement}) to be of some 80\,m/s. This includes the noise level. The $1\sigma$-value of the 'straight' mode is $1\sigma=67$\,m/s if no mask is applied, which is of the same order of magnitude as the VTF sensitivity. For magnetic pixels (last two columns in Tab.~\ref{tab:vlosmaps}), the $1\sigma$-error is more than double the size of the noise level: $1\sigma=182$\,m/s for $B_{\rm los}>100$\,G and  201\,m/s for $B_{\rm los}>200$\,G. Hence, the $1\sigma$-errors from solar evolution are substantial for a measurement span of 11\,s.

In the 'normalised straight' mode, the $1\sigma$-error decreases substantially by roughly a factor of two: $1\sigma=36$\,m/s, if no mask is applied, 88\,m/s for $B_{\rm los}>100$\,G and 94\,m/s for $B_{\rm los}>200$\,G. The errors of the rabbit mode are slightly better than the straight mode, but not as small as for the normalised straight mode.

Comparing the errors of 's1-s2' and 's1-s11' demonstrates that the temporal evolution is significant: The differences between the first two snapshots are small, much smaller than the VTF sensitivity.  They are probably mostly caused by the low spectral resolution which does not allow for a more accurate determination. However, the errors between the first and the eleventh snapshot increase by a factor of eight or more: from $1\sigma=18$\,m/s  to 195\,m/s for all pixel, and from $1\sigma=50$\,m/s  to 433\,m/s for the pixels with $B_{\rm los}>100$\,G.

We also computed the histograms for spatially binned data and list the results in Tab.~\ref{tab:binning} for the magnetic pixels with $B_{\rm los}$ between 0.1 and 2.5\,kG. We find that the $1\sigma$-errors of the straight mode decreases from $1\sigma=181$\,m/s for no binning, to $1\sigma=169$\,m/s for two by two binning, to $1\sigma=148$\,m/s for four by four binning. Hence, even with a four by four binning the time-evolution error exceeds significantly the VTF sensitivity if the measurement is done in the straight mode. A four by four binning would correspond to the diffraction limit of a 1m-aperture telescope.

Note that in case, the measurement objective is to measure only velocities and no magnetic fields, the non-polarimetric Doppler mode can be chosen in which one accumulation suffices to measure the Stokes-$I$ line profile. This measurement only takes 0.9\,s (VTF IPC, v.3.4, see Sect.~\ref{sec:measurement}), and deviations due to solar evolution can be safely neglected.

\subsubsection{Errors in $B_{\rm  los}$ maps}

In Tab.~\ref{tab:Bvermaps} the error values characterising the distribution of the $\triangle B_{\rm los}$ map are listed. These errors are to be compared to the VTF measurement sensitivity of 20\,G, which results from the photon noise (see Sect.~\ref{sec:measurement}). It is seen that the $1\sigma$ errors for the entire map and for weak-field pixels ($<100$\,G) are negligible for straight and normalised straight modes. 

However, within the magnetic filigree the time-evolution errors are substantial. Considering all pixels with $B_{\rm los}$ between 100 and 2500\,G, the error amount to 48\,G in the straight mode and to 36\,G in the normalised straight mode, which is significantly larger than the 20\,G error due to noise. Interestingly, the rabbit mode has an error of 84\,G and thus performs worse than both the straight and the normalised straight mode. It seems that the Milne-Eddington inversion can deal better with profiles, in which neighbouring wavelength points are recorded close in time. It is not straightforward to explain this behaviour.

The differences between two consecutive snapshots 's1-s2' are negligible and smaller than 20\,G in all cases. The time evolution is well reflected in the differences between the first and the eleventh snapshot, 's1-s11'. These differences correspond to an error of $1\sigma=91$\,G for the magnetic pixels exceeding 100\,G.

Hence, for filigree and magnetic features which are shuffled around in the inter-granular lanes, the fields are strong enough, and more accumulation and longer effective exposure times will not improve the measurement. Due to their dynamic behaviour and strong fields, it may even be better to measure them with less accumulations, i.e, shorter effective exposure times. For measuring magnetic fields in and above granules, the situation is different: the evolution time scale is longer and the magnetic fields are weaker. Hence, for granular magnetic fields more accumulations might be wanted to improve the signal-to-noise ratio. However, one should be aware that a homogeneous horizontal field of 100\,G produces a $Q$-amplitude of only some 0.0017 for Fe I \,617.3\,nm.

\subsection{Implication for data pipelines}\label{sec:pipeline}

We have seen that the errors due to solar evolution can in all cases be reduced by normalising the continuum intensity to the instanteous continuum intensity during the measurement as indicated by Eq.~(\ref{eq:norm}). This can be done in our case, because we compute the whole line profile and its continuum for each snapshot. However, a Fabry-Perot instrument measures, of course, only one wavelength point at a time. 

But instruments like VTF, CRISP, CHROMIS, \& IBIS simultaneously measure a broad-band image at a neighbouring wavelength. These broad-band images can be used to mimic the temporal variation of the continuum intensity. To our knowledge, this is not currently done in existing Fabry-Perot data pipelines. 

In order to apply Eq.~(\ref{eq:norm}), one needs to scale the broad-band images to the continuum intensity of narrow-band images. As seen in Fig.~\ref{fig:convolution}, a complication may exist, if a true continuum wavelength point is not part of the measurement sequence. Hence, in practise two approaches could do the job: (1) Add one more wavelength point to include a true continuum point outside the line. This point can then be used to determine the scaling factor between the broad-band intensity and the narrow-band continuum intensity. (2) In case the continuum intensity cannot be measured, e.g., because the line is too broad for the narrow pre-filter\footnote{The pre-filter is needed to suppress side peaks from the etalon(s).}, one would compute the average measured profile of a quiet Sun region, fit a quiet Sun profile to it, and determine the continuum intensity from the fitted profile to scale the broad-band intensities.

\section{Conclusions}\label{sec:conclusions}

We investigate the measurement errors that arise due to temporal changes during the measurement process. Already with the 1.5\,m telescope GREGOR (see Fig.~\ref{fig:gregor}), it can be seen that magnetic features in inter-granular lanes are shuffled around on a time scale of 11\,s. This is also seen in MURAM numerical simulations of a region with a mean vertical field strength of 200\,G (see Fig.~\ref{fig:cont_diff}). Based on these simulations, we mimic the measurement process of VTF and determine the measurement error by defining the 'truth' to be the temporal average of the time that elapses during the measurement. 

We find that granules are less affected and inter-granular areas with magnetic features are strongly affected. Depending on the science objective, it might be advisable to adapt the effective exposure time in the measurement: For measuring strong fields in the inter-granular lane only a few accumulations may suffice, and reduce the measurement error as the effective exposure time is reduced. For weaker fields within granules, more accumulations can be afforded as the evolution time scale is longer. And for predominantly horizontal (perpendicular to line-of-sight) fields, more accumulations are necessary, because the amplitude of linear polarisation in horizontal fields is much weaker than circular polarisation in vertical fields.

A key result of this investigation is the finding, that a normalisation of the narrow-band intensities with the temporally averaged continuum intensity, following Eq.~(\ref{eq:norm}) (and see Fig.~\ref{fig:composition} for an illustration), leads to a significant reduction of the measurement errors. In Sect.~\ref{sec:pipeline}, we discuss how this normalisation could be integrated in existing pipelines, taking advantage of the simultaneously measured broad-band image.

\begin{acknowledgements}
We thank Matthias Rempel for providing the MURAM simulations, and Markus Schmassmann for making the data accessible and software to read the data. We thank Philip Lindner for providing the GREGOR data used in Section 1. 
\end{acknowledgements}

\bibliographystyle{aa}
\bibliography{vtf}

\end{document}